\documentclass{aa} % for a referee version
\usepackage{graphicx,natbib}
\usepackage{txfonts}
\begin{document}
%
%\title{Resolving the Rotation-Activity Connection in Very Cool
%  Stars\thanks{Based on observations carried out at the European
%    Southern Observatory, La Silla, PID~076.D-0092}}

\title{The Narrowest M-dwarf Line Profiles and the Rotation-Activity
  Connection at Very Slow Rotation\thanks{Based on observations
    carried out at the European Southern Observatory, La Silla,
    PID~076.D-0092}}

%\subtitle{Observations of the Narrowest Line Profiles in M-dwarfs}

   \author{A. Reiners\inst{1,2}\thanks{Marie Curie Outgoing International Fellow}}

   \offprints{A. Reiners}

   \institute{Universit\"at G\"ottingen, Institut f\"ur Astrophysik, Friedrich-Hund-Platz 1, D-37077 G\"ottingen, Germany\\
   \email{Ansgar.Reiners@phys.uni-goettingen.de}
   \and 
   Universit\"at Hamburg, Hamburger Sternwarte, Gojenbergsweg 112, D-21029 Hamburg, Germany
  }

   \date{Received 21 December 2006 / Accepted 21 February 2007 }

% \abstract{}{}{}{}{} 
% 5 {} token are mandatory
 
  \abstract
  {The rotation-activity connection explains stellar activity in terms
    of rotation and convective overturn time. It is well established
    in stars of spectral types F--K as well as in M-type stars of
    young clusters, in which rotation is still very rapid even among
    M-dwarfs. The rotation-activity connection is not established in
    field M-dwarfs, because they rotate very slowly, and detecting
    rotation periods or rotational line broadening is a challenge.  In
    field M-dwarfs, saturation sets in below $v_\mathrm{rot} =
    5$\,km\,s$^{-1}$, hence they are expected to populate the
    non-saturated part of the rotation-activity connection.}
  {This work for the first time shows intrinsically resolved spectral
    lines of slowly rotating M-dwarfs and makes a first comparison to
    estimates of convective velocities. By measuring rotation
    velocities in a sample of mostly inactive M-dwarfs, the
    unsaturated part of the rotation-activity connection is followed
    into the regime of very low activity.}
  {Spectra of ten M-dwarfs are taken at a resolving power of $R =
    200\,000$ at the CES in the near infrared region where molecular
    FeH has strong absorption bands. The intrinsically very narrow
    lines are compared to model calculations of convective flows, and
    rotational broadening is measured.}
  {In one star, an upper limit of $v\,\sin{i} = 1$\,km\,s$^{-1}$ was
    found, significant rotation was detected in the other nine
    objects. All inactive stars show rotation below or equal to
    2\,km\,s$^{-1}$.  In the two active stars AD\,Leo and YZ\,CMi,
    rotation velocities are found to be 40--50\,\% below the results
    from earlier studies.}
  { The rotation activity connection holds in field early-M stars,
    too.  Activity and rotation velocities of the sample stars are
    well in agreement with the relation found in earlier and younger
    stars. The intrinsic absorption profiles of molecular FeH lines
    are consistent with calculations from atomic Fe lines.
    Investigation of FeH line profiles is a very promising tool to
    measure convection patterns at the surfaces of M-stars.}


   \maketitle
%
%________________________________________________________________

\section{Introduction}

The connection between rotation and activity among sun-like stars
allows a look inside the nature of the magnetic dynamo hosted in
sun-like stars. The rotation-activity connection is established in
late-type stars (mid F -- early M) through all ages and spectral types
for which rotation can be measured \citep[e.g.,][]{Noyes84, Patten96}.
It is believed that a rotation-dependent dynamo process generates
large-scale magnetic fields that heat the upper layers of the
atmosphere and cause the variety of activity phenomena observed in the
Sun and other stars. Tracers of stellar activity are X-ray emission
and chromospheric emission lines like H$\alpha$ and Ca~H \& K. For a
given spectral type, the strength of activity depends on rotation in
the sense that among slow rotators (here, the definition of ``slow''
depends on spectral type, see below) activity grows with rotation
speed, but at a certain angular velocity, activity becomes saturated
and does not grow with more rapid rotation.

Between different spectral types, the dependence of activity on
stellar rotation rate is slightly different. This is believed to be
due to the different efficiencies of the stellar dynamo, which depends
on the dynamo number $N_\mathrm{D}$ \citep{Parker71}, but there is
also some debate whether this dependence is only seen when normalizing
emission measures to bolometric luminosity \citep{Basri86,
  Pizzolato03}. \cite{Durney78} suggested a relationship between
dynamo number $N_\mathrm{D}$ and Rossby number $Ro$ of the form
$N_\mathrm{D} \approx Ro^{-2}$ \citep[see also][]{Noyes84}, while $Ro
\propto P/\tau_\mathrm{conv}$, with $P$ the rotation period and
$\tau_\mathrm{conv}$ the convective overturn time, the latter is the
factor that depends on spectral type.  The criterion for a dynamo to
work is $N_\mathrm{D} \ga 1$ ($Ro \la 1$), i.e., rotation is stronger
than convection. \cite{Noyes84} showed that plotting the chromospheric
emission ratio $R'_\mathrm{HK}$ vs.  Rossby number significantly
reduces the scatter that appears when plotting $R'_\mathrm{HK}$ vs.
rotation period.  \cite{Patten96, Randich96} and \cite{Pizzolato03}
used this to show that the rotation-activity connection is universal in
cluster and field stars of different age and spectral type.

\subsection{Rotation and activity in early M-stars}

\begin{figure}
  \includegraphics[width=.5\textwidth]{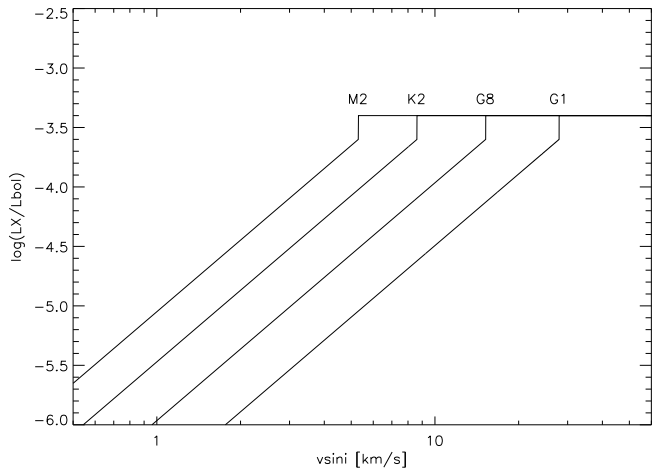}
  \caption{\label{fig:LX_vsini_expect}Mean rotation activity
    connection from \cite{Patten96} with Rossby number translated into
    surface velocity shown for four spectral types. Due to different
    radii and convective turnover times, the relation spreads in
    spectral type, in the M-dwarfs the rising part falls below $v =
    5$\,km\,s$^{-1}$ }
\end{figure}

Stars of spectral type M are places of frequent flaring events. These
stars have relatively low temperatures while flaring activity stems
from high temperature plasma, thus their UV and optical colors are
subject to enormous brightness variations. Activity in field M-stars
was observed in X-rays \citep[e.g.,][]{Barbera93, Fleming93,
  Schmitt95, Giampapa96} and H$\alpha$ \citep[e.g.,][]{Herbst89,
  Hawley96, Delfosse98, Gizis02, Mohanty03, West04}. The Ca~H \& K
lines are difficult to measure since they have very little flux in the
M-stars.  All measurements led to the picture that M-stars can be very
active, and \cite{West04} shows that the fraction of active M-stars is
rather low at early subclasses but rises dramatically around spectral
type M4, up to $\sim 75$\,\% at M7 before it goes down again at the
late M-type objects. The falloff towards spectral class L is explained
by the high electrical resistivity in such cool atmospheres
\citep{Fleming00, Mohanty02}.

The validity of the rotation-activity connection was shown in a large
range of stars of different age and spectral type. Measurements of
activity are available for a huge number of targets provided by a
wealth of different instruments either spectroscopically, from
dedicated imaging projects or large sky surveys. Measuring rotation,
however, requires either to follow brightness variations over some
time to observe periodic changes due to presumably corotating features
on the stellar surface, or the analysis of a high-quality spectrum to
detect rotational line broadening in spectral features.

The saturated part of the rotation-activity connection, i.e. fast
rotators, is well established since short periods as well as strong
rotation broadening are relatively easy to measure. The non-saturated
part of this connection, however, consists of the slower rotators with
longer rotation periods and weaker rotational line broadening. This
poses a big problem to the measurement of the rotation-activity
connection in M-stars, where convective overturn times are longest
leading to rotation periods of several weeks and -- due to the small
radii -- to very small surface velocities.

The measurement of rotation periods in M-stars is particularly
difficult since on active stars flares occur on timescales much
shorter than the rotation period which effectively hides the rotation
period under the noise of stellar activity. The measurement of a long
period requires extremely high data quality and time coverage
\citep[e.g.][]{Benedict98}. If, on the other hand, a star is inactive,
there is little rotational variation in its lightcurve so that the
detection of a period is difficult or impossible. Detections of
rotation periods in M-stars are still very limited; \cite{Pettersen82}
reports rotation periods in 11 Me-dwarfs in the range 1--8\,d, all of
them translate into surface velocities larger than 4\,km\,s$^{-1}$.
\cite{Torres82} report three significant detections of rotation
periods, again in dMe stars. The amount of questionable results
(indicated with questionmarks in their Table\,II) and the paucity of
later successful detections of M-star rotation periods demonstrate
the difficulty detecting them\footnote{Note that \cite{SH86} claim the
  detection of a rotation period in AD~Leo that is comparable to the
  one of \cite{Torres82} (indicated as questionable in their work).
  The fact that both groups find the same period may suggest that
  AD~Leo indeed rotates at $P \approx 2.7$\,d.  However, the detection
  remains somewhat tentative in both works.}.

To follow the rotation activity connection to the slowest rotators in
old (hence slowly rotating) M-stars, the measurement of the rotation
velocity is often the easier way. How does the rotation activity
relation look if one uses $v\,\sin{i}$ instead of rotation period or
Rossby number? In Fig.\,\ref{fig:LX_vsini_expect}, the mean rotation
activity connection from \cite{Patten96} is shown as an activity vs.
equatorial velocity plot. Due to different radii and convective
turnover times, the one relation found for spectral types mid-F to
early-M spreads in spectral type. In the low-mass stars, the rising
part is shifted to smaller velocities. This is shown for four
different spectral types, G1, G8, K2, and M2 in
Fig.\,\ref{fig:LX_vsini_expect}. While stars of spectral type G1 are
saturated at a velocity of almost 30\,km\,s$^{-1}$, M2-stars are
already saturated at about 5\,km\,s$^{-1}$. Note that the translation
of the empirical relation also depends on the chosen value for the
overturn time $\tau_\mathrm{conv}$, which is difficult to compute
particularly in very cool stars. The semi-empirical values of
\cite{Noyes84} were chosen for Fig.\,\ref{fig:LX_vsini_expect}.
\cite{Gilliland86} and \cite{Kim96} also calculated
$\tau_\mathrm{conv}$ and gave somewhat higher $\tau_\mathrm{conv}$ for
M-stars. This would translate into even smaller rotation velocities
for the M2 example.

In Fig.\,\ref{fig:LX_vsini_expect}, one can see that resolving the
rotation activity connection among field M-dwarfs requires the
measurement of rotational line broadening on the order of
1\,km\,s$^{-1}$.  Unfortunately, two mechanisms work against the
observer; a) high spectral resolution requires a lot of photons, but
the targets are very faint; b) detecting slow rotation requires sharp
spectral features, but spectra of M-dwarfs are covered with molecular
absorption bands eating away the continuum.  Individual atomic lines
as in the solar spectrum are hardly available at all since most lines
are too weak at such low temperatures, and the strong alkali lines are
so pressure broadened that the signature of slow rotation is washed
out.

\subsection{Measuring very slow rotation}

The measurement of Doppler line broadening due to very slow rotation
requires high spectral resolution so that it is not dominated by
instrument effects. In principle, one can correct for instrumental
broadening if accurately known. In the presence of noise and blended
spectral lines, however, rotational broadening is only detectable if
it is on the order of the instrument profile. In other words, slow
rotation on the order of 1\,km\,s$^{-1}$ has very little influence on
a line profile observed at a limited resolution of, e.g.,
5\,km\,s$^{-1}$, but it will be easily detectable if the instrumental
profile is only 1\,km\,s$^{-1}$ wide.  Furthermore, broadening effects
other than rotation (convection, Zeeman splitting, etc.) may influence
the line shape as well. The line profile of a ``non-rotating'' M-star
has a FWHM on the order of 6-7\,km\,s$^{-1}$ at a resolution of $R
\sim 50\,000$ and on the order of 2\,km\,s$^{-1}$ at a resolution of
$R \sim 200\,000$. With the lower resolution, such a profile is
sampled in only 1--2 resolution elements while at high resolution it
is seen in about 5--10 resolution elements. Clearly, projected
rotation velocities of $v\,\sin{i} \approx 2$\,km\,s$^{-1}$ are
difficult to detect in the former case but easily detectable in the
latter.

Another technique that is frequently used to determine rotation in
M-dwarfs is the cross-correlation technique. The width of the
cross-correlation profile between a slowly rotating template and the
target star is used as an indicator for rotation velocity. This
strategy partly overcomes the problem of low SNR and the lack of
isolated spectral features since it utilizes larger wavelength regions
that show characteristic features. Molecular bands have been used even
in late M- and L-dwarfs \citep{Mohanty03, Bailer04}.  However, one
cannot expect the method to yield lower detection thresholds at very
low rotation since the width of the cross-correlation profile cannot
be more sensitive to small rotation than the direct comparison of
narrow lines is (note that the cross-correlation profile is generally
broader than the narrowest lines, this weakens the advantage of the
higher SNR of the cross-correlation profile at slow rotation).
Furthermore, even small template mismatch and other systematic
uncertainties (e.g., uncertainties in the continuum normalization)
always lead to a \emph{wider} cross-correlation profile, i.e. too high
a rotation velocity -- no effect can lead to a cross-correlation
profile that is narrower than the one for the correct rotation
broadening.

In this paper, I present the first spectra of absorption lines that
resolve intrinsic line broadening in M stars, i.e. isolated absorption
lines not dominated by the instrumental profile or rotation. Molecular
absorption of FeH around $1\,\mu$m are intrinsically narrow and well
isolated so that they provide an ideal tracer of M-dwarf line
broadening due to convection or very slow rotation \citep[$v\,\sin{i}
< 3$\,km\,s$^{-1}$, ][]{Reiners06a}. At a resolving power of $R \sim
200\,000$, these data carry information on slow rotation and surface
velocity patters that are not contained in data of lower spectral
resolution. The data will be described in more detail in
Section\,\ref{sect:data}. Rotation velocities are measured and put
into context with activity measurements in
Sections\,\ref{sect:analysis} and \ref{sect:results}. A summary is
provided in Section\,\ref{sect:summary}.

\section{Data}
\label{sect:data}

\begin{figure*}
  \centering
  \includegraphics[width=.9\textwidth,bbllx=0,bblly=0,bburx=594,bbury=290]{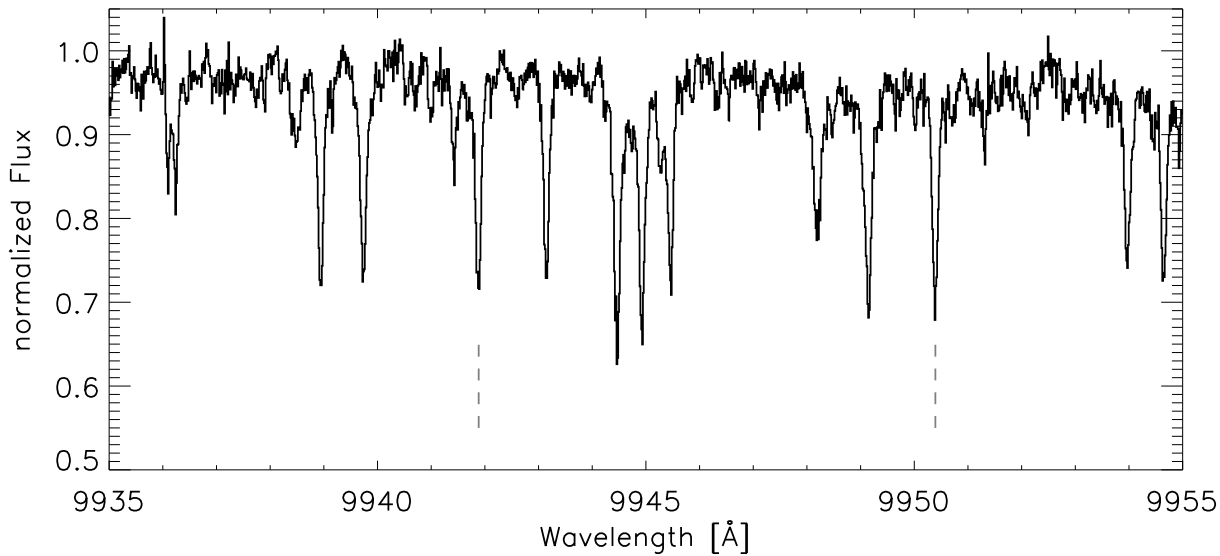}
  \caption{\label{fig:region}Part of the spectrum of Gl\,273 observed
    with the CES. The two magnetically insensitive lines at
    9941.9\,\AA\ and 9950.4\,\AA\ that are used for ration analysis
    are marked.}
\end{figure*}

\begin{table}
\begin{minipage}[t]{\columnwidth}
  \caption[]{\label{tab:Objects} Sample stars, observations and targets
    used for flatfielding}
  \renewcommand{\footnoterule}{}  % to avoid a line before footnotes
      \begin{tabular}{lcccc}
        \hline
        \hline
        \noalign{\smallskip}
        Name & SpType\footnote{\cite{Reid95, Hawley96}} & $m_\mathrm{I}$\footnote{\cite{Leggett92}} & exp. time & SNR\footnote{per pixel, SNR per resolution element is about a factor of two higher}\\
        &&& [min] &\\
        \noalign{\smallskip}
        \hline
        \noalign{\smallskip}
        Gl 514   &  M0.5 & 7.04 & 65 & 45 \\ 
        Gl 229A  &  M1.0 & 6.11 & 60 & 80 \\
        Gl 526   &  M1.5 & 6.43 & 45 & 45 \\
        Gl 205   &  M1.5 & 5.88 & 50 & 85 \\
        Gl 382   &  M1.5 & 7.09 & 85 & 50 \\
        Gl 393   &  M2.0 & 7.41 & 60 & 45 \\
        Gl 273   &  M3.5 & 7.16 & 110 & 50 \\
        AD Leo   &  M3.5 & 6.81 & 60 & 40 \\
        Gl 628   &  M3.5 & 7.40 & 30 & 30 \\
        YZ CMi   &  M4.5 & 8.20 & 90 & 20 \\
        \noalign{\medskip}
        \multicolumn{5}{c}{Telluric Standards used as Flatfields}\\
        \noalign{\smallskip}
        HD 74956 & A1V & 1.86 & 10 & \\
        HD 47670 & B8III & 3.24 & 30 & \\
        \noalign{\smallskip}
        \hline
      \end{tabular}
\end{minipage}
\end{table}     

Data were taken at the Coud\'e Echelle Spectrograph (CES) at the 3.6m
telescope, ESO, La Silla. The spectral region at 1\,$\mu$m is
exceptionally red for its front-illuminated CCD resulting in
relatively poor efficiency and heavy fringing. However, since the
spectral region is near the peak of the spectral energy distribution
of M-stars, high signal-to-noise ratios (SNR) could be achieved with
relatively short exposures. Exposure times and SNR are given in
Table\,\ref{tab:Objects} together with the apparent I-magnitudes from
\cite{Leggett92}. The high resolution of $R \sim 200\,000$ is achieved
through an image slicer chopping the light from the star in 12 narrow
slices. The instrument is fibre-fed and no geometric information is
conserved after the light has passed through fibre and slicer.
Unfortunately, this means that the fringing pattern of the CCD depends
on how the fibre is illuminated. As a consequence, the fringing in the
on-star observations cannot be corrected for using a flatfield
exposure that uniformly illuminates the fibre instead of providing a
point source as the stars do. Thus, the flatfields taken at the CES
are useless in this spectral region.  In order to correct for the
heavy fringing, high-SNR exposures of two telluric standard stars were
taken, and they were used as 2D flatfields during the reduction
procedure.  Observations of both flatfields were combined to yield
highest SNR.\footnote{Flatfielding was done in the 2D-images.  The
  reduced flatfield spectrum would have a SNR of $\sim 300$.}

Using telluric reference stars to correct for the pixel to pixel
sensitivity also eliminates telluric features if observations of the
target stars are taken at very similar airmass. As shown in
\cite{Reiners06a}, the spectral region near 1$\mu$m is virtually free
of telluric contamination so that no telluric correction seems
necessary. Thus, standard stars were taken at arbitrary position
(furthermore, no stars bright enough for this project are available
next to each target).

The sample consists of the brightest slowly rotating M-dwarfs visible
in early spring at La Silla, Chile. Several inactive stars with
rotation velocities too low for previous measurements were chosen in
order to resolve slow rotation in the unsaturated part of the
rotation-activity connection at spectral types where stars become
fully convective.  Three stars with detected rotation were observed as
well, two of them (AD~Leo and YZ~Cmi) very active stars with saturated
coronal emission.

\section{Rotation analysis}
\label{sect:analysis}

In this paper, rotation is measured comparing the shape between lines
of the slowest rotator (the template) and the other targets. Rotation
velocities are found by artificially broadening the narrow lines of
the template star. To do so, it is essential to find isolated and
intrinsically narrow spectral lines that are embedded in a clear
continuum. In the classical optical wavelength regions, such lines are
difficult to find since TiO and VO have strong absorption systems in
this region.  However, at longer wavelengths around 1$\mu$m, the main
opacity sources TiO and VO do not show absorption bands, and the only
features are lines of molecular FeH. This absorption band was
investigated by \cite{Reiners06a}, it consists of many isolated lines
of FeH, many clearly discernible and intrinsically narrow.  The lines
are embedded in a clear continuum which allows to accurately follow
individual line profiles from the core to the wings. Among the M
dwarfs, the structure of the FeH band follows an optical depth scaling
so that FeH lines of any M dwarf spectrum can be reproduced by scaling
the FeH band of a template star that may have a different spectral
class \citep{Reiners06a}. A part of the CES spectrum in the FeH region
is shown in Fig.\,\ref{fig:region}.  The two lines, $\lambda 9941.9$
and $\lambda 9950.4$\AA, that are used to determine rotational
broadening are indicated in the figure.

The most reliable way determining rotational line broadening is to
utilize the spectrum of a star that shows virtually no rotation, i.e.
$v\,\sin{i}$ is below the detection limit. The width of the narrowest
line that can possibly be observed in a spectrum depends I) on the
instrumental resolution, and II) on the intrinsic line width without
rotation, i.e. effects like temperature broadening, pressure
broadening, and convection. The star with the narrowest profile in the
sample is Gl\,273, which is also the star with the lowest ratio
$L_{\mathrm X}/L_\mathrm{bol}$. No rotation has been detected in this
star before and there is no reason to assume that significant
broadening due to rotation affects its line profile. Nevertheless, the
line profiles of the FeH lines in Gl\,273 are well resolved and they
are significantly broader than the instrumental profile as discussed
in the following.

Recently, \cite{LM07} has shown that a connection exists between
radius and metallicity in single M dwarfs, and between radius and
activity in very active members of binary systems.  The stars in the
sample analyzed here are single stars with relatively low activity, so
that no effect regarding the radius of the stars is expected through
strong activity.  Metallicity, however, is not known for many of the
objects, so that the radii of the stars could differ among the same
spectral type.  However, the FeH lines analyzed in this work are
insensitive to gravity (and hence radius).  No effects on the
derivation of $v\,\sin{i}$ are expected.

\subsection{The slowest rotator}

\begin{figure*}
  \mbox{
    \includegraphics[width=.5\textwidth]{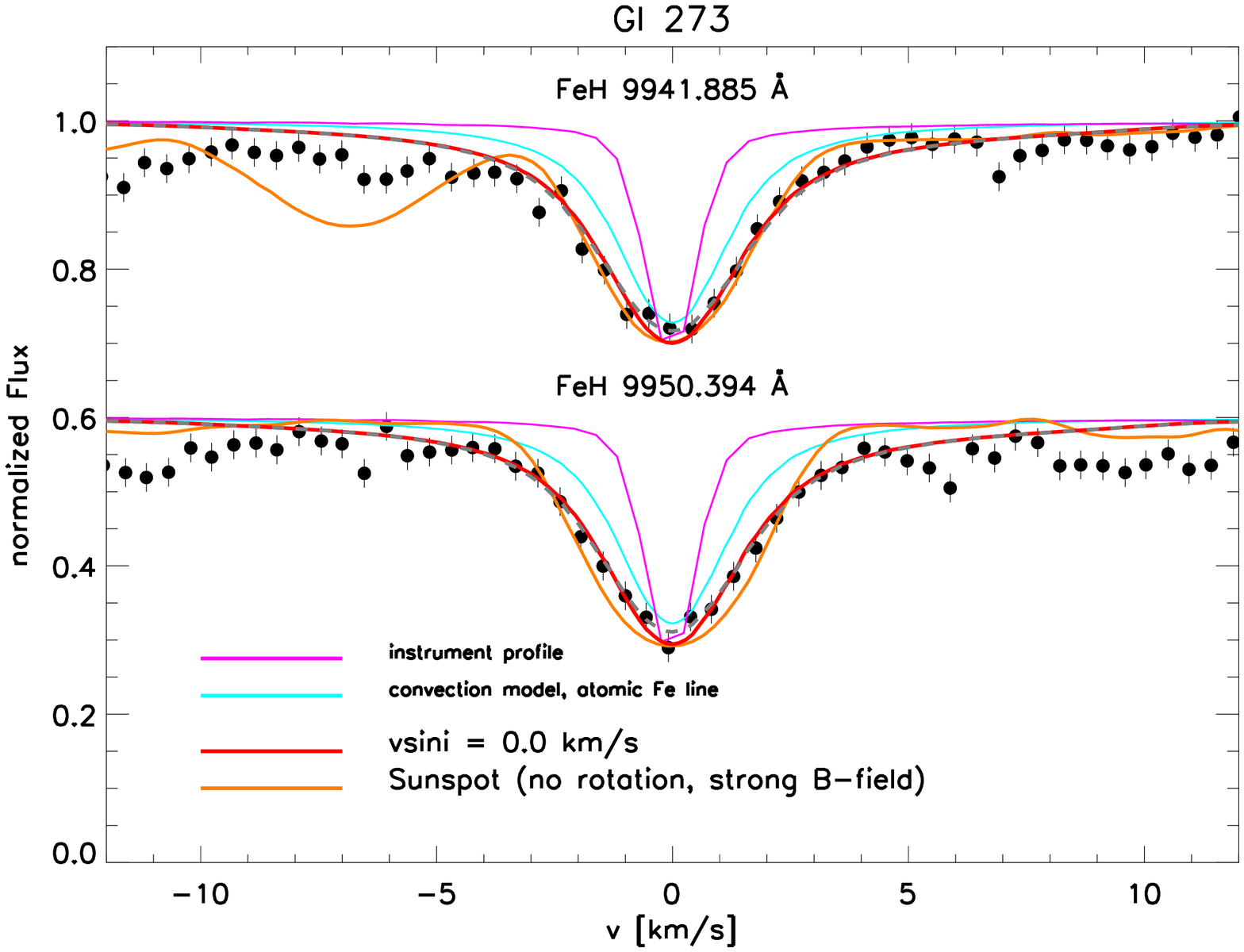}
    \includegraphics[width=.5\textwidth]{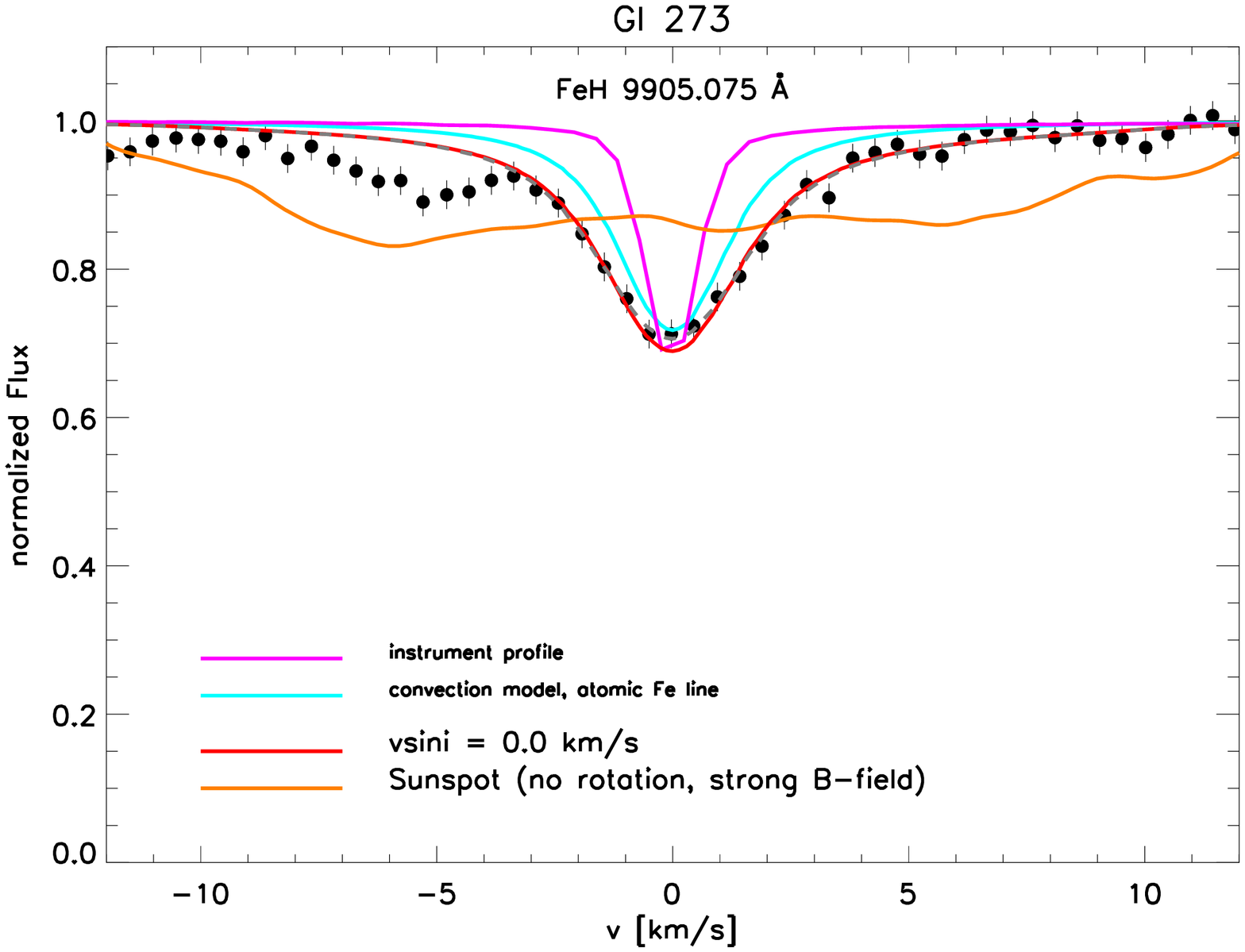}
  }
  \caption{\label{fig:Gl273}Spectral lines of Gl\,273. Overplotted are
    the instrumental profile (magenta), a model of an Fe~I line
    (cyan), the convolution of instrumental and model profiles (red),
    and a sunspot spectrum of the same region (orange).  \emph{Left:}
    The two magnetically insensitive lines; \emph{right:} a
    magnetically sensitive line, note the strong Zeeman broadening in
    the sunspot spectrum.}
\end{figure*}

The two spectral lines $\lambda 9941.9$ and $\lambda 9950.4$\,\AA\ of
Gl\,273 are shown in the left column of Fig.\,\ref{fig:Gl273} in
velocity units relative to the line center. Data are plotted with
error bars according to the measured SNR given in
Table\,\ref{tab:Objects}. The instrumental profile measured from ThAr
lines is overplotted (magenta), it is significantly smaller than the observed
line showing that the line shape has been resolved.  Unfortunately,
radiative transfer calculations of the FeH lines are not yet available
so that it was not possible to compare the line shape to a fully
consistent model of the two molecular lines.  However, the main
broadening mechanisms in such narrow lines are temperature broadening
and convective velocities. The atomic weight of Fe is a good
approximation of the weight of the FeH molecule, thus temperature
broadening in Fe lines can be expected to be very similar to the
effect in FeH lines. The strength of the FeH lines in early M stars is
also comparable to some Fe lines, so that the line formation depths
should also be similar in both species.  Thus, line broadening due to
convective motions are probably of comparable strength.

In Fig.\,\ref{fig:Gl273}, a model of the intrinsic profile of the Fe~I
line at $\lambda = 6518.37$\,\AA\ for $T_\mathrm{eff} = 2800$\,K and
$\log{g} = 5.0$ is shown (cyan). The profile was kindly provided by
\cite{Sven06}.  Atmospheric parameters correspond to an M6 atmosphere
which is slightly cooler than our target objects. It is not the aim of
this study to perform a detailed comparison to the model spectrum,
which only serves as a consistency check for the observed line profile
in the slowest rotator. The difference in temperature broadening
between the model star and Gl\,273 implies a difference between the
observed lines and the model line of $\sim 60$\,m/s, i.e. the model
spectrum is expected to be about 60\,m/s narrower that the one of
Gl\,273. Convective velocities have not yet been observed in M-dwarfs,
but they are not expected to change significantly within the early-
and mid-M stars either.\footnote{The vanishing of a radiative core
  does not significantly affect the photospheric velocities.} Thus,
there is no reason to believe that the intrinsic spectrum of Gl\,273
(i.e., the spectrum of an early M star with no rotation) is smaller
than the modeled spectral line.

The convolution of the instrumental profile and the synthetic Fe~I
line is overplotted over the data in Fig.\,\ref{fig:Gl273} (red line).
The Fe~I line was scaled using an optical depth scaling to account for
saturation (the scaling factor was approximately unity).  It is
immediately clear that our expectation of an observed Fe~I line has a
shape that very accurately matches the line shape observed in the two
FeH lines in question.

What does the consistency of synthetic Fe~I lines and the FeH lines
mean in terms of rotational broadening? The intrinsic line broadening
in Fe and FeH lines can be expected to be similar since the main
broadening mechanisms are comparable; in particular there is no reason
to assume that the FeH lines in Gl~273 should be \emph{narrower} than
the Fe~I lines at the atmospheric parameters used in the model. That
means that no indication for extra broadening due to rotation is found
in the observed lines. The similarity of the two FeH lines in Gl~273
to the shape expected for observed Fe lines means that within the
uncertainties the FeH lines are consistent with zero rotation.

\subsection{Zeeman sensitivity of the spectral lines}

In Fig.\,\ref{fig:Gl273}, a spectrum of a sunspot\footnote{Available
  at ftp://ftp.noao.edu/fts/spot3atl} is plotted over the data
(orange). This spectrum is formed in a somewhat hotter environment
than the M dwarf spectrum and in a strong magnetic field ($\sim
2$\,kG), but it is not rotationally broadened since the spot is
spatially resolved.  The two lines shown in the left panel resemble
the two lines in the spectral region that are least affected by Zeeman
broadening consistent with the analysis of the FeH band observed at
lower resolution in \cite{Reiners06a}. In the first line ($\lambda
9942$\,\AA), the sunspot spectrum resembles the M-dwarf spectrum very
accurately, it is a little broader in the core (probably due to the
larger temperature broadening) and an extra component shows up at
$-7$\,km\,s$^{-1}$.  Again, there is no indication for rotational
broadening in the integrated spectrum of Gl\,273 from comparing its
FeH lines to the sunspot spectrum. The second line ($\lambda
9950$\,\AA) appears somewhat broader in the sunspot spectrum.  This
cannot be ascribed to convection since it is not seen in the $\lambda
9942$\,\AA\ line (where the sunspot line matches the M dwarf line),
hence the extra broadening in the FeH line at $9942$\,\AA\ must be due
to the strong magnetic field of the sunspot.  The strong effect of
Zeeman broadening on a magnetically sensitive FeH line is shown in the
right column of Fig.\,\ref{fig:Gl273}, in which the same as in the
left panel is shown but for the line at $\lambda = 9905$\,\AA.  This
line is very sensitive to magnetic fields, which can be seen in the
heavy broadening of the sunspot spectrum. This comparison also shows
that Zeeman broadening is extremely small in the two lines of the left
panel. Thus, for the analysis of rotational broadening, we can
minimize the effect of Zeeman broadening employing only the two
magnetically insensitive lines (left panel in Fig.\,\ref{fig:Gl273}).
As a byproduct, it becomes clear that Gl~273 can only have a very weak
mean magnetic field (that the product of magnetic field and surface
coverage is very small).

In this paper I will not determine the magnetic fluxes of the observed
targets. \cite{Reiners06a} show a method to measure the magnetic flux
interpolating between the spectrum of a star with strong flux and a
star with weak or zero flux. With the CES data, a more accurate
analysis of the strength and of the pattern of Zeeman broadening in
lines of the FeH molecule is possible but goes beyond the scope of
this paper.

\subsection{Detection threshold}

\begin{figure*}
  % \begin{minipage}[t]{\textwidth}
  \centering
  \mbox{
    \includegraphics[width=.5\textwidth,bbllx=0,bblly=0,bburx=648,bbury=468]{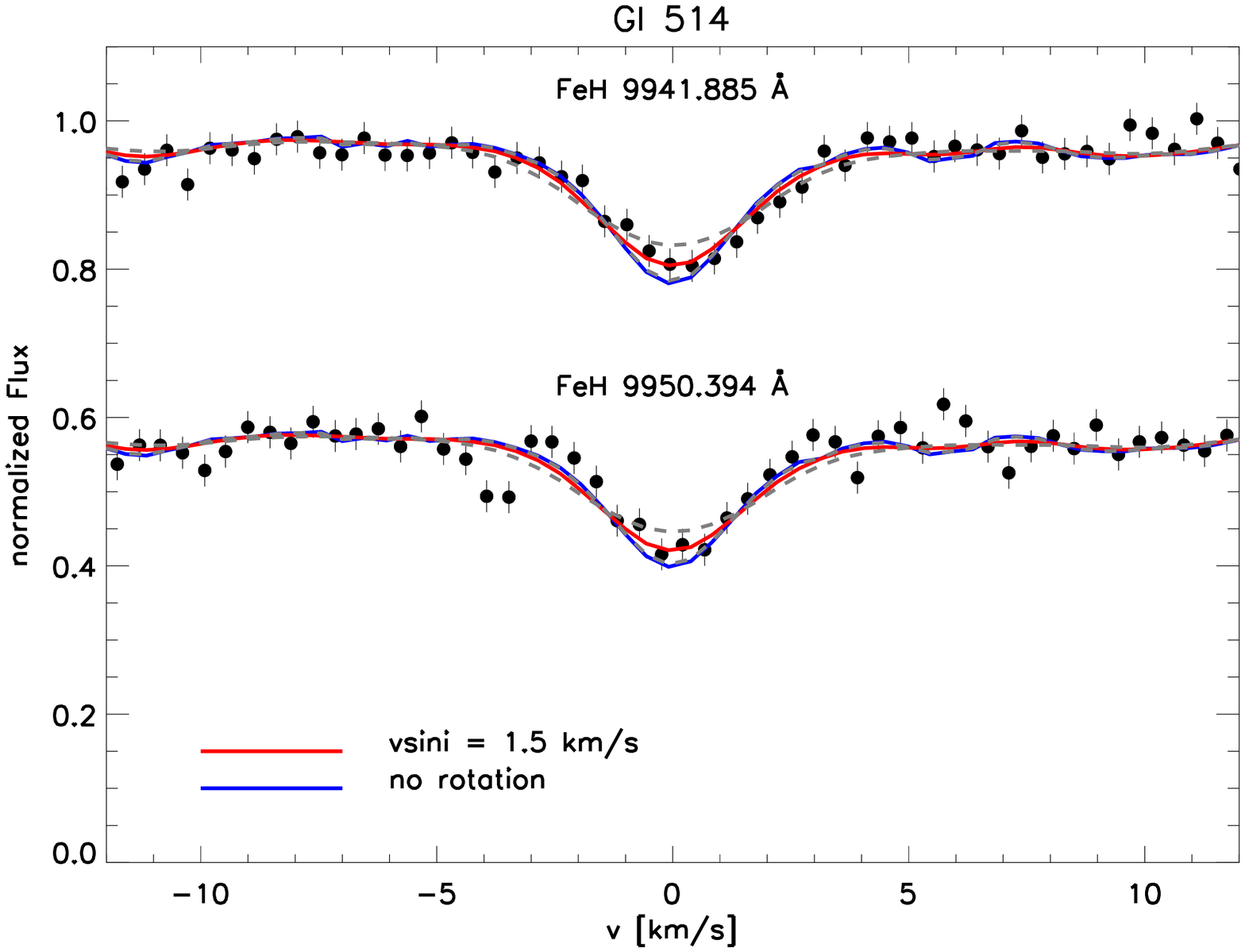}
    \includegraphics[width=.5\textwidth,bbllx=0,bblly=0,bburx=648,bbury=468]{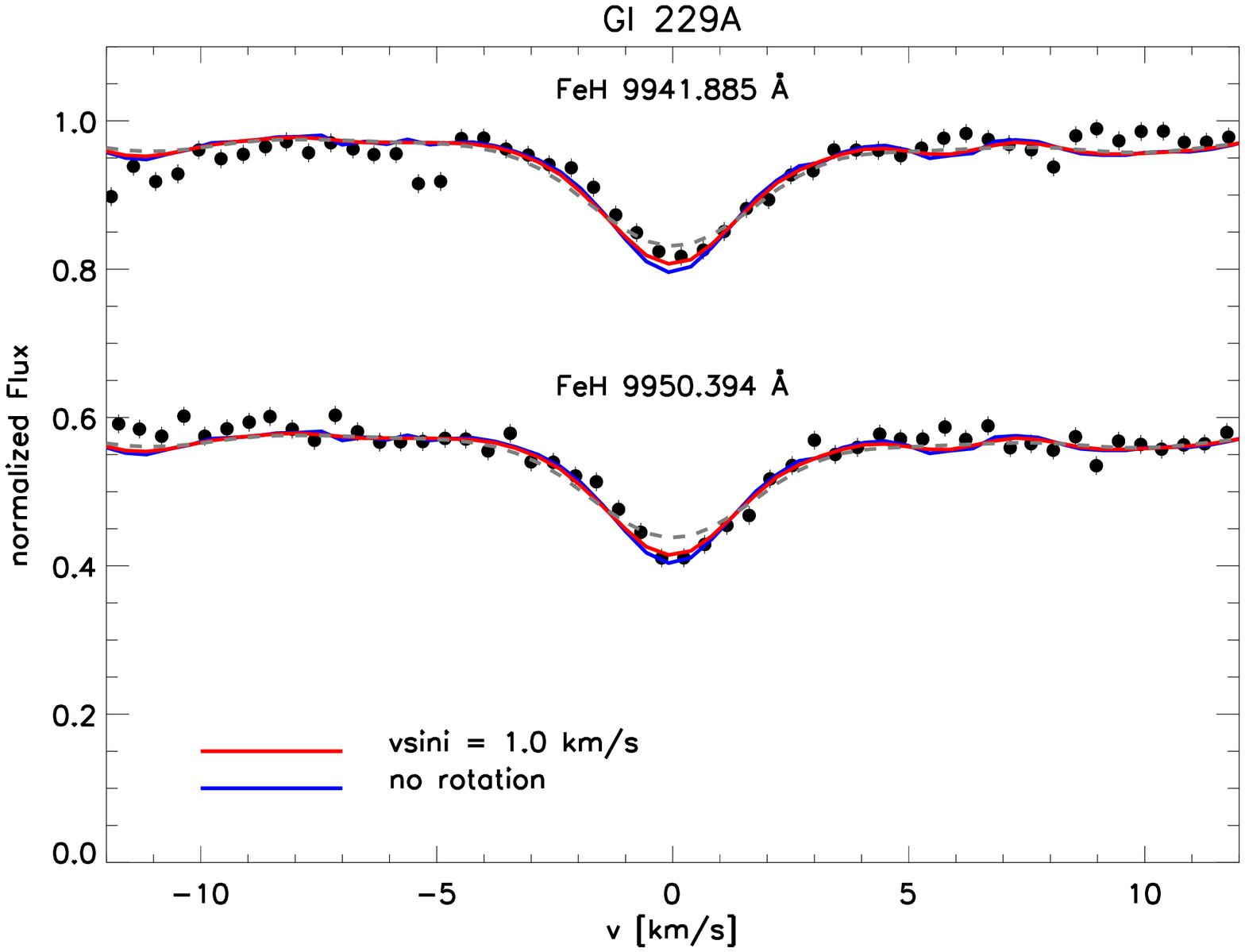}
  }
  \caption{\label{fig:Fits1}The two magnetically insensitive spectral
    lines of Gl~514 (left) and Gl~229A (right). The template profile
    (see text) is overplotted in blue, the red line shows the best fit
    with rotational broadening given in the plot. Dashed lines show
    artificial broadening for $v\,\sin{i} \pm 1$\,km\,s$^{-1}$.}
\end{figure*}

\begin{figure*}
  \centering
  \mbox{
    \includegraphics[width=.5\textwidth,bbllx=0,bblly=0,bburx=648,bbury=468]{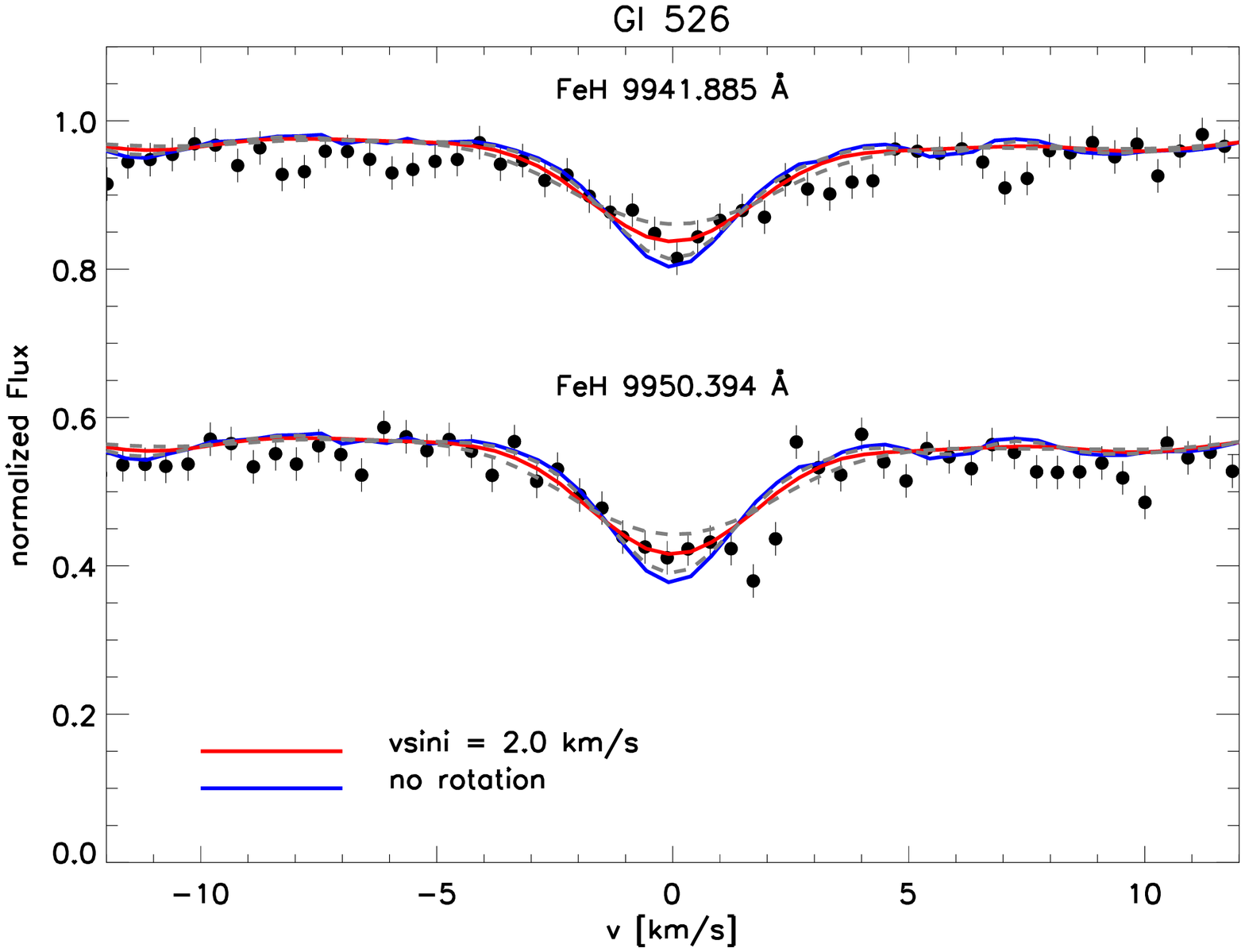}
    \includegraphics[width=.5\textwidth,bbllx=0,bblly=0,bburx=648,bbury=468]{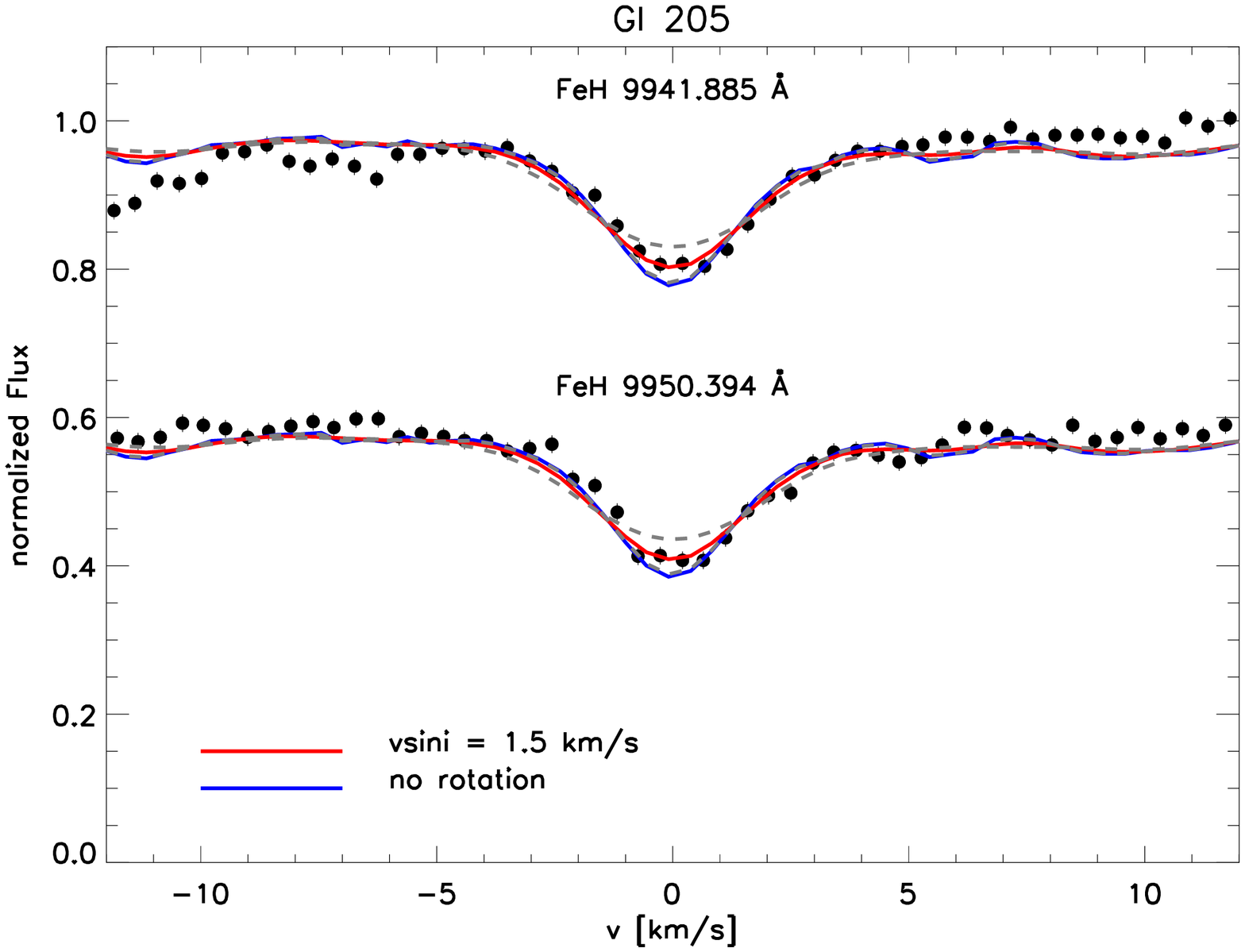}
  }
  \caption{\label{fig:Fits2}Same as Fig.\,\ref{fig:Fits1} but for Gl~526 (left) and Gl~205 (right).}
\end{figure*}

\begin{figure*}
  \centering
  \mbox{
    \includegraphics[width=.5\textwidth,bbllx=0,bblly=0,bburx=648,bbury=468]{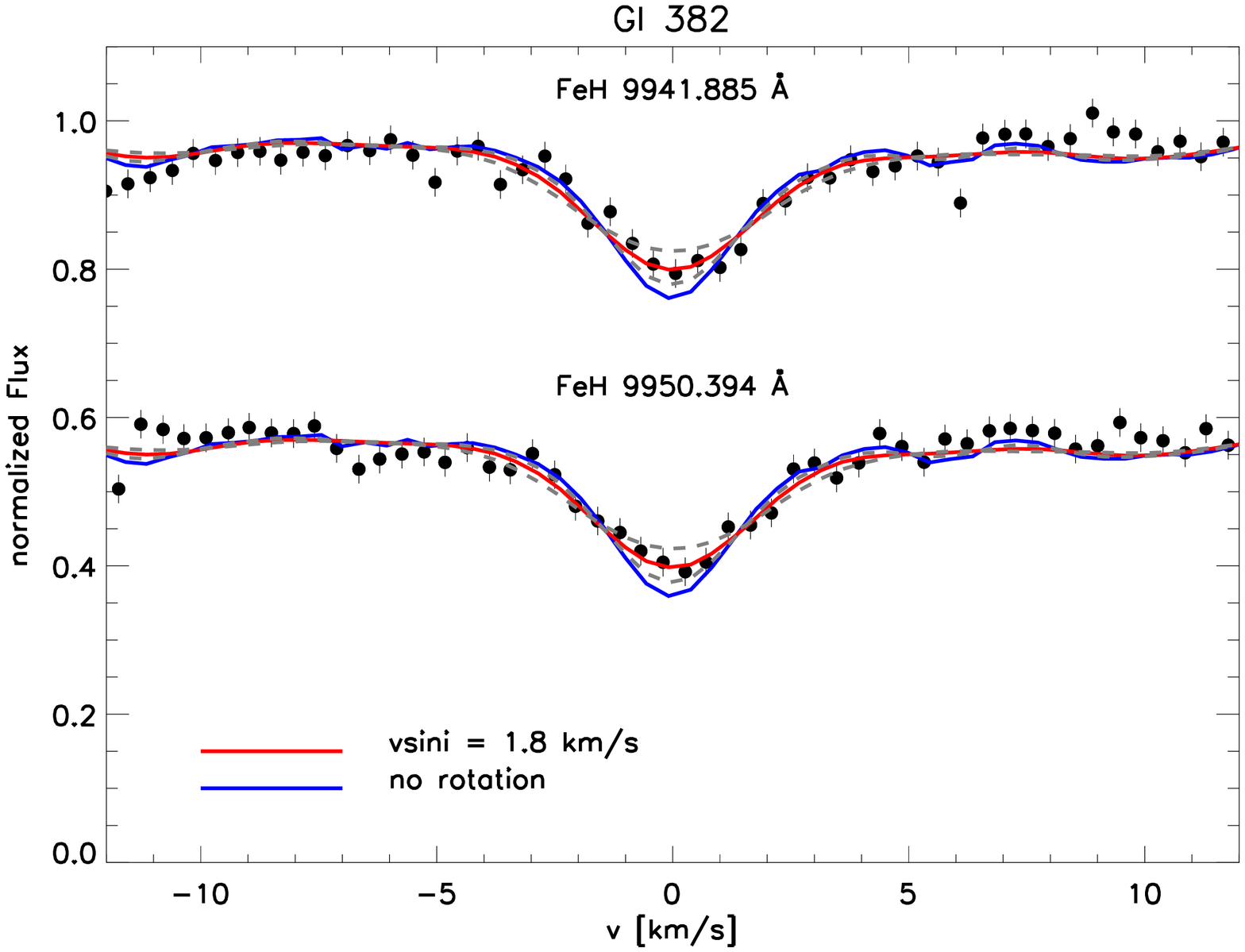}
    \includegraphics[width=.5\textwidth,bbllx=0,bblly=0,bburx=648,bbury=468]{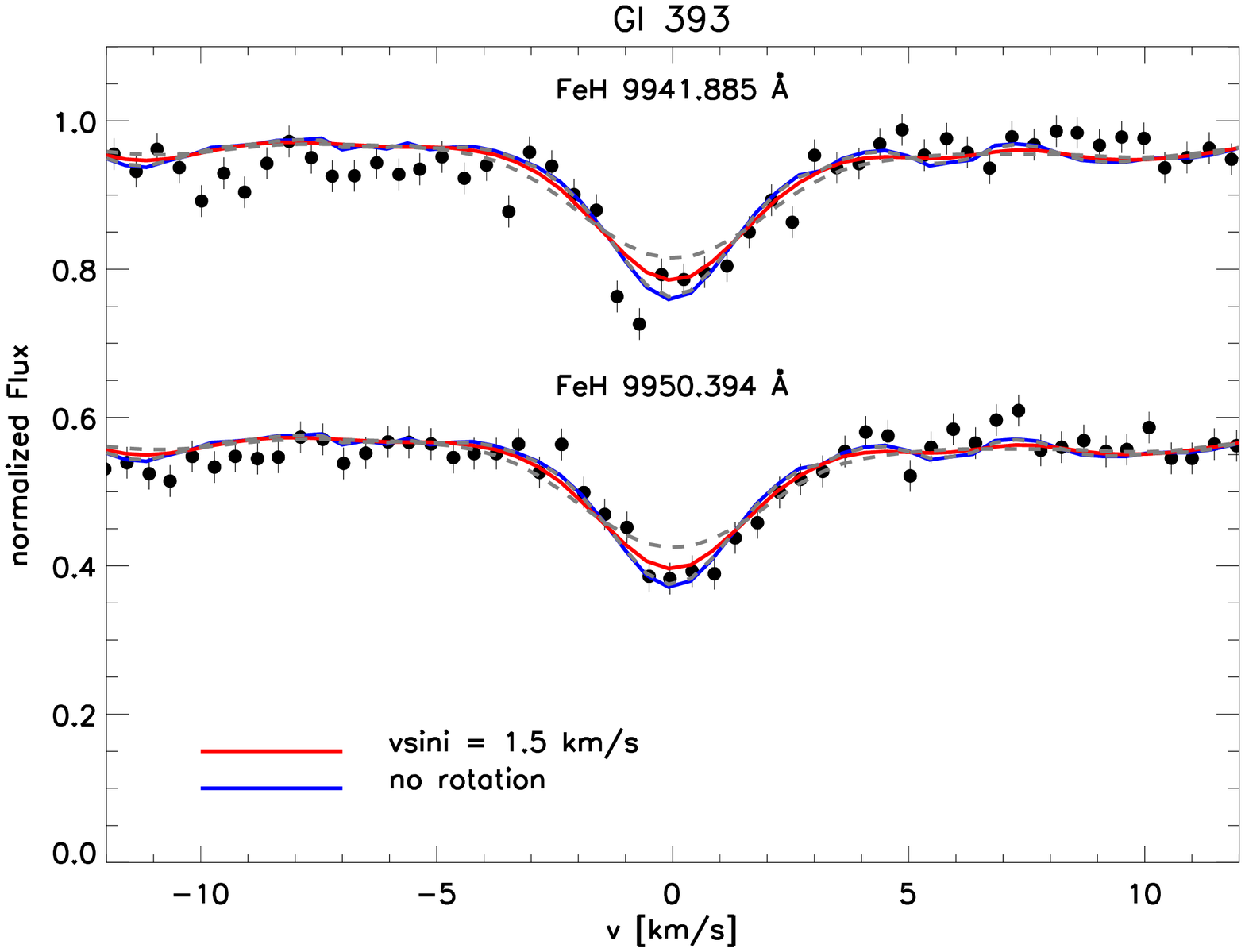}
  }
  \caption{\label{fig:Fits3}Same as Fig.\,\ref{fig:Fits1} but for Gl~382 (left) and Gl~393 (right).}
  % \end{minipage}  
\end{figure*}

\begin{figure*}
  \centering
  \mbox{
    \includegraphics[width=.5\textwidth,bbllx=0,bblly=0,bburx=648,bbury=468]{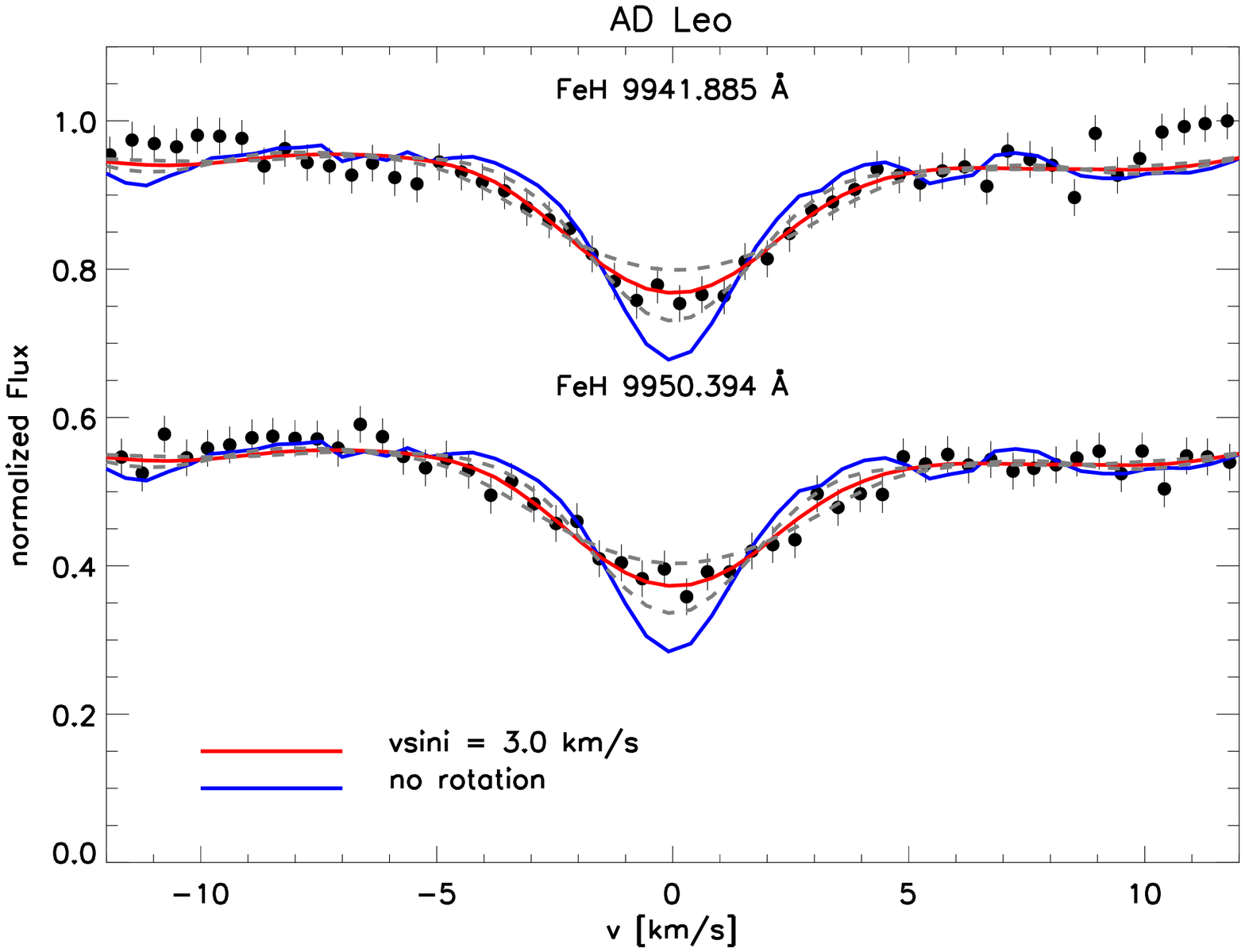}
    \includegraphics[width=.5\textwidth,bbllx=0,bblly=0,bburx=648,bbury=468]{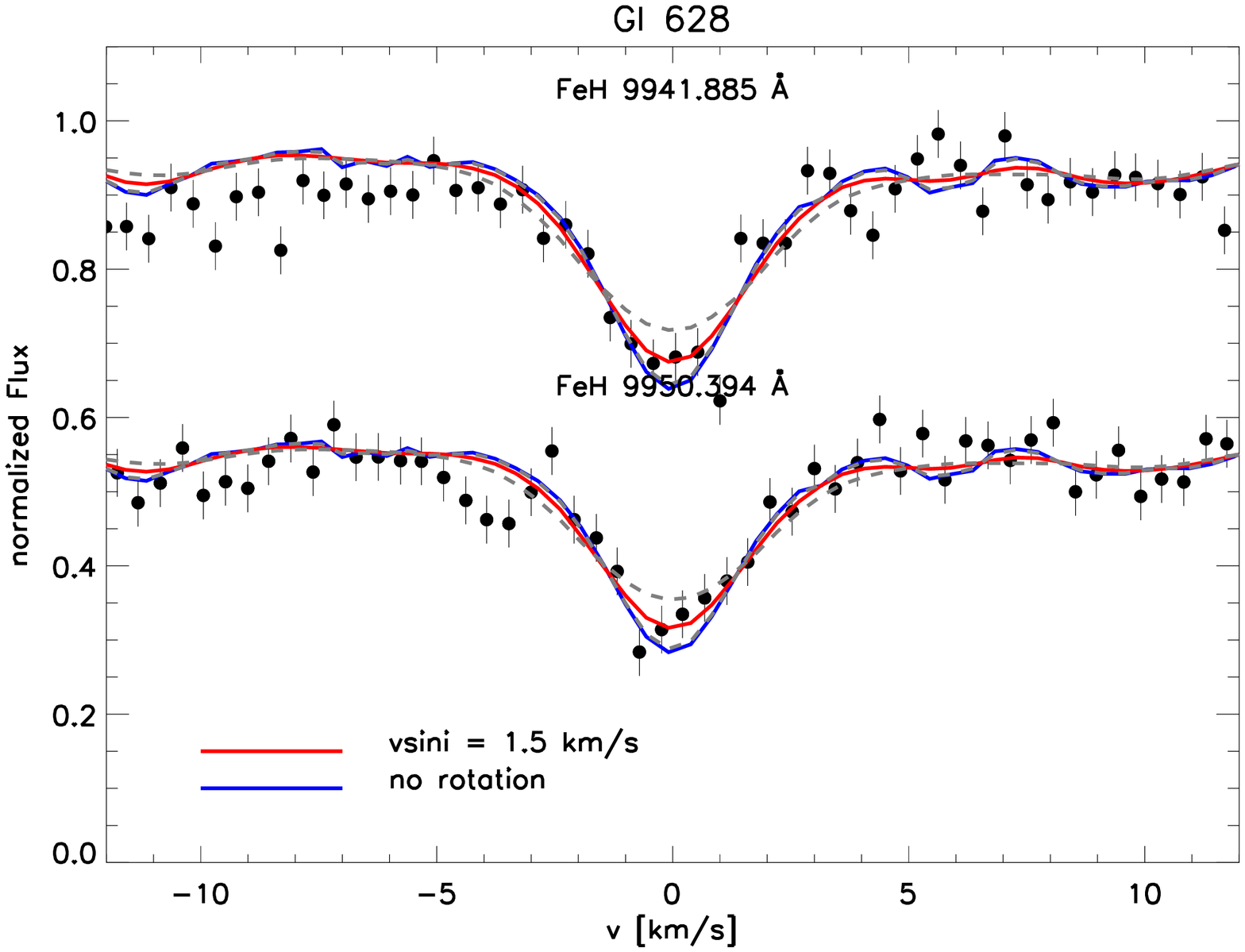}
  }
\caption{\label{fig:Fits4}Same as Fig.\,\ref{fig:Fits1} but for AD~Leo (left) and Gl~628 (right).}
\end{figure*}

To estimate the lowest rotation velocity that can be detected in our
data, the comparison spectrum, i.e. the Fe~I line broadened according
to the instrument resolution, was artificially spun up to a
$v\,\sin{i}$ of 1\,km\,s$^{-1}$. The result is overplotted in
Fig.\,\ref{fig:Gl273} as dashed lines, but it essentially remains
hidden below the curve for $v\,\sin{i} = 0$\,km\,s$^{-1}$. Such small
a rotation broadening is difficult to see even in the high quality
data used here. It is important to realize that a rotation velocity of
$v\,\sin{i} = 1$\,km\,s$^{-1}$ is hardly detectable at a resolution of
$R = 200,000$ in an M-star, which has a very narrow intrinsic line
shape due to the small temperature broadening. At lower resolving
power this threshold grows, and small rotation becomes more difficult
to detect.

The comparison shows that a projected rotation velocity of $v\,\sin{i}
= 1$\,\,km\,s$^{-1}$ cannot be ruled out for Gl\,273.  At higher
velocities, however, a difference of 1\,km\,s$^{-1}$ is rather easy to
detect (compare to the case of Gl\,229A in the right panel of
Fig.\,\ref{fig:Fits1}) as soon as rotational broadening becomes a
significant fraction of the intrinsic width. Thus, from visual
comparison, the detection threshold for rotation broadening in data of
this quality is estimated to $\Delta v\,\sin{i} = 1$\,km\,s$^{-1}$, and
$v\,\sin{i} \le 1$\,km\,s$^{-1}$ is adopted for Gl\,273.

\subsection{Significant rotation}

All other sample stars show FeH lines significantly broader than the
lines in Gl\,273. Since both lines are not sensitive to Zeeman
broadening, the extra broadening must be due to rotation or
differences in convective motion. The latter is probably very small in
the narrow range of spectral types (Gl\,273 has a spectral type of
M\,3.5), and I will assume that rotation is the only extra broadening
mechanism for what follows. To determine the value of $v\,\sin{i}$,
the template spectrum (see below) is artificially broadened searching
for the best match to the data in both lines. The best fit was
determined by eye, which is sufficient to the level of accuracy
achieved here (see below). In all cases, the two FeH lines yielded
consistent results.

The template lines were constructed using a combination of the
artificial spectrum of the Fe~I line and the observed lines from
Gl\,273. In order to minimize the effects of limited SNR in the
spectrum of Gl\,273, that spectrum was smoothed using a 3 pixel
box-car. The smoothed spectrum was only used in the wings of the
template spectrum. In the core of the lines, such an approach would
lead to an effective broadening of the line mimicking higher rotation
velocity. On the other hand, the artificial spectrum of the broadened
Fe~I line accurately resembles the line core of the two lines, and it
was used instead of the original data spectrum in the core region
between $\pm3$\,km\,s$^{-1}$.

\begin{figure}
  \centering
  \mbox{
    \includegraphics[width=.5\textwidth,bbllx=0,bblly=0,bburx=648,bbury=468]{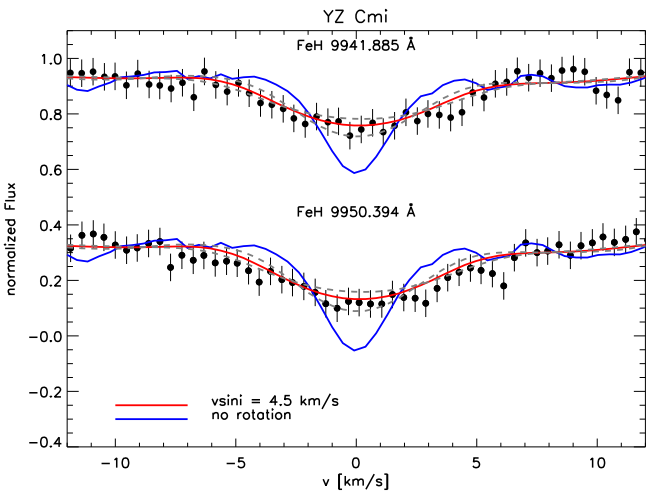}
  }
\caption{\label{fig:Fits5}Same as Fig.\,\ref{fig:Fits1} but for YZ~Cmi.}
\end{figure}

The template spectrum was scaled and artificially broadened to match
the spectra of the sample targets. Figs.\,\ref{fig:Fits1} --
\ref{fig:Fits5} show the two FeH lines for all stars together with the
best fits. For comparison, two artificial spectra at $v\,\sin{i} =
v\sin{i}_{\mathrm{fit}} \pm 1$\,km\,s$^{-1}$ are also plotted (dashed
lines) demonstrating the sensitivity in $v\,\sin{i}$.  It can be seen
that the data quality allows to determine the value of $v\,\sin{i}$
within an uncertainty of $\pm 1$\,km\,s$^{-1}$.

\subsection{Rotation velocities from cross-correlation}
\label{sect:rotvel}

A very successful method used to measure stellar rotation velocities
in stars and brown dwarfs of spectral types M and L is the cross
correlation method \cite[e.g.,][]{Basri00, Mohanty03, Bailer04}.  A
spectrum of a slowly rotating star is used as a template, it is
artificially broadened and the width of the cross-correlation function
between the unbroadened template and its broadened version is
calibrated for various rotation speeds. Then, the cross-correlation
between a star and the template is calculated and the width of this
profile is compared to the set of calibrations obtained before. One
caveat of this method is that it cannot account for Zeeman broadening.
Effectively, an average of the rotation broadening from all lines is
calculated.  Hence it could be argued that for the magnetically active
stars the cross-correlation method yields systematically too high
values for $v\,\sin{i}$. This hypothesis can be tested by comparing
results from cross-correlation to results from direct line fitting of
magnetically insensitive lines.

I calculated the cross-correlation functions for the sample stars in
the full wavelength range of the observations using the slowest
rotator in the sample, Gl~273, as a template.  A second set of
$v\,\sin{i}$ values is calculated this way, the results are given in
the fourth column of Table\,\ref{tab:Results}. It is immediately clear
that in the domain of very slow rotation, $v\,\sin{i} \la
5$\,km\,s$^{-1}$, the result from line fitting and cross-correlation
agree very well, less (if any) influence can be expected at higher
rotation. It appears that magnetic broadening does not influence the
measurement of rotation via the cross-correlation method in this
spectral region. The difference in spectral type does not appear to
influence this result. It is important to note that this spectral
region is dominated by absorption from FeH lines, and that the
structure of this band does not significantly change in the spectral
range investigated. In a region with different absorbing species (and
larger spectral coverage) larger differences can be expected.

\section{Results}
\label{sect:results}
\subsection{Rotation of the sample stars}

\begin{table*}
\begin{minipage}[t]{\textwidth}
  \caption{\label{tab:Results} Results}
  \renewcommand{\footnoterule}{}  % to avoid a line before footnotes
  \begin{tabular}{lccccccrcccr}
    \hline
    \hline
    \noalign{\smallskip}
    Name & $B-V$\footnote{\citet{Leggett92}} & $v \sin{i}$ & $v \sin{i}_{\mathrm{Xcorr}}$ & $v \sin{i}_{\mathrm{Lit}}$ & & $R$\footnote{\citet{Lacy77}} & $\frac{P}{\sin{i}}$& $L_\mathrm{bol}$\footnote{Using calibration from \citet{Delfosse98}, M$_{V}$ from \citet{Gliese91} and $R-I$ from \citet{Leggett92}} & $L_\mathrm{X}$\footnote{From NEXXUS database \citep{NEXXUS}, i.e. ROSAT data.} & log($\frac{L_\mathrm{X}}{L_\mathrm{bol}}$) & $\log{Ro}$ \\
    && km\,s$^{-1}$ & km\,s$^{-1}$ & km\,s$^{-1}$ &&R$_\odot$& d\hspace{1.5mm} & erg s$^{-1}$ & erg s$^{-1}$\\
    \noalign{\smallskip}
    \hline
    \noalign{\smallskip}
    Gl 514   &  1.49 & 1.5 & 1.3 & $<2.9$\footnote{\cite{Delfosse98}}    &&  0.60 & 20.3    & 32.11 & 27.38 & -4.73 &   $-$0.11\\ 
    Gl 229A  &  1.50 & 1.0 & 1.0 & $\hspace{2.5mm}3.0$\footnote{\cite{Vogt83}} &&  0.69 & 35.0    & 32.27 & 27.11 & -5.16 &   0.11\\
    Gl 526   &  1.43 & 2.0 & 1.4 & $<2.9^{e}$    &&  0.50 & 12.7    & 32.12 & 26.87 & -5.25 &   $-$0.32\\
    Gl 205   &  1.47 & 1.5 & 1.0 & $<2.9^{e}$    &&  0.79 & 26.8    & 32.39 & 27.66 & -4.73 &   0.00\\
    Gl 382   &  1.50 & 1.8 & 1.3 & $<2.9^{e}$    &&  0.69 & 19.4    & 32.26 & 27.45 & -4.81 &   $-$0.14\\
    Gl 393   &  1.51 & 1.5 & 1.1 & $<2.9^{e}$    &&  0.52 & 17.7    & 32.02 & 26.98 & -5.04 &   $-$0.19\\
    Gl 273   &  1.57 &$\le$1.0 & 0.0\footnote{Used as template for Cross Correlation} & $<2.4^{e}$    &&  0.35 & $>$17.9     & 31.64 & 26.03 & -5.61 &   $-$0.19\\
    AD Leo   &  1.53 & 3.0 & 3.0 & $\hspace{2.5mm}6.2^{e}$ &&  0.50 & 8.4 & 31.94 & 28.58 & -3.36 &   $-$0.51\\
    Gl 628   &  1.58 & 1.5 & 1.1 & $<1.1^{e}$    &&  0.32 & 10.7 & 31.60 & 26.79 & -4.81 &   $-$0.42\\
    YZ CMi   &  1.61 & 4.5 & 5.3 & $\hspace{2.5mm}6.5^{e}$ &&  0.30 & 3.4 & 31.68 & 28.57 & -3.11 &   $-$0.92\\
    \noalign{\smallskip}
    \hline
  \end{tabular}
\end{minipage}
\end{table*}

In all stars but one, significant rotational line broadening is
detected. Only the spectral lines of Gl\,273 do not show any
broadening in excess of what is expected from the Fe~I model and the
instrumental profile. The spectral lines of Gl\,273 are also
significantly narrower than the lines in all other stars. The
uncertainty and the detection threshold for the projected rotation
velocity $v\,\sin{i}$ is $\sim 1$\,km\,s$^{-1}$ (see above).

The measured values of $v\,\sin{i}$ are given in
Table\,\ref{tab:Results} together with literature values from former
publications. In all but one cases, this work yields rotation
velocities lower than what was previously reported. In six stars,
literature contained only upper limits of about 3\,km\,s$^{-1}$ for
$v\,\sin{i}$, and this work provides detection of rotation of or below
2.0\,km\,s$^{-1}$ for five of them. In Gl\,273, \cite{Delfosse98}
report an upper limit of $v\,\sin{i} \le 2.4$\,km\,s$^{-1}$ (although
this accuracy seems a little optimistic given the resolution of their
data). The upper limit on the projected rotation velocity of Gl\,273
is now as low as $v\,\sin{i} \le 1.0$\,km\,s$^{-1}$. In three stars, a
detection of rotation was reported before, the three cases are the
following:

\begin{enumerate}
\item Gl 229A ($\mathbf{v\,sin{i} = 1.0}$\,km\,s$\mathbf{^{-1}}$):
  \cite{Vogt83} report $v\,\sin{i} = 3.0$\,km\,s$^{-1}$, but in their
  table of results they indicate that this value is uncertain. They
  used a Ba~II line at $\lambda = 6141$\,\AA\ at a resolution of $R =
  50\,000$ (FWHM of 6\,km\,s$^{-1}$).  Such low a rotation velocity
  has a very subtle effect in a single line and their result may be
  interpreted as an upper limit as well.
\item AD Leo ($\mathbf{v\,sin{i} = 3.0}$\,km\,s$\mathbf{^{-1}}$): This
  famous flare star has been a frequent subject to rotation
  measurements. \cite{Vogt83, Marcy92, Delfosse98, Fuhrmeister04} all
  measure $v\,\sin{i}$ between 5 and 7.6 km\,s$^{-1}$ from either fits
  to single atomic lines \citep{Vogt83, Marcy92}, the cross
  correlation method \citep{Delfosse98}, or fitting large parts of the
  spectrum \citep{Fuhrmeister04}. However, they all use wavelength
  regions that do not contain isolated lines but either single lines
  embedded in a continuum affected by molecular absorption, or
  molecular bands that consist of many blended features. The
  resolution of the data they use is also a factor of 2--4 lower than
  what was available here.  Apparently, the higher data quality and in
  particular the use of isolated features allows the detection of the
  low rotation velocity in AD~Leo while slow rotation is hidden in
  lower resolution data in broad parts of the spectrum.
  \cite{Reiners06b} also determine $v\,\sin{i} \approx
  3$\,km\,s$^{-1}$ from data in the same wavelength region at lower
  spectral resolution and explicitly taking into account Zeeman
  broadening.  Whether effects of magnetic fields mimic more rapid
  rotation at other wavelength regions, is difficult to assess.
  AD~Leo is a flare star and \cite{Reiners06b} measure a net magnetic
  flux of $Bf \sim 3$\,kG, thus significant broadening may affect the
  spectral features in large parts of the spectrum.  However,
  \cite{Marcy92} addressed this question finding that the lines they
  used are not affected by Zeeman broadening, and the results from
  cross-correlation analysis in Section~\ref{sect:rotvel} show that
  Zeeman broadening does not lead to a systematic offset in the
  measurements of $v\,\sin{i}$ even in a wavelength region that
  contains absorption features with rather strong Zeeman sensitivity.
\item YZ Cmi ($\mathbf{v\,sin{i} = 4.5}$\,km\,s$\mathbf{^{-1}}$):
  \cite{Delfosse98} and \cite{Fuhrmeister04} find a value of
  $v\,\sin{i}$ of 6.5\,km\,s$^{-1}$ instead of the 4.5\,km\,s$^{-1}$
  reported here.  \cite{Marcy92} find $v\,\sin{i} = 4.8$\,km\,s$^{-1}$
  but admit poor data quality. YZ~CMi is also a flare star with an
  even stronger magnetic flux \citep[$Bf \sim 4$\,kG according
  to][]{Reiners06b}, hence the case of YZ~Cmi may be similar to
  AD~Leo. However, the difference of 2\,km\,s$^{-1}$ may not be
  considered significant, and the new (lower) value comes from
  significantly better data.
\end{enumerate}

\subsection{The rotation-activity connection in very slowly rotating M-stars}

\begin{figure}
  \includegraphics[width=.5\textwidth]{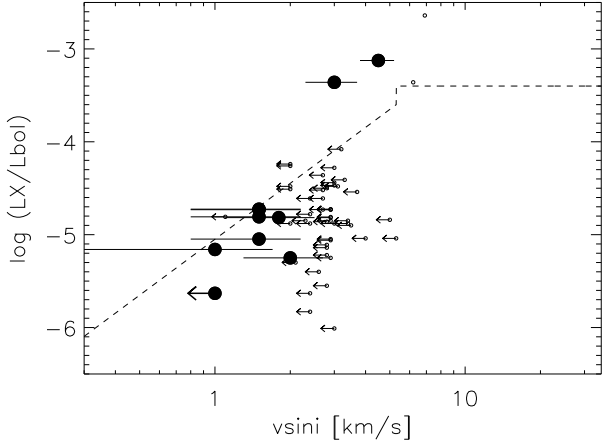}
  \caption{\label{fig:LX_vsini} $\log{L_\mathrm{X}/L_\mathrm{bol}}$
    vs.  $v\,\sin{i}$ for early M-dwarfs (including YZ~Cmi, M4.5). New
    rotation measurements are plotted as full circles, data from
    \cite{Delfosse98} as small open circles (most of their
    $v\,\sin{i}$ values are upper limits).
    $\log{L_\mathrm{X}/L_\mathrm{bol}}$ for the stars of this sample
    were recalculated as explained in the text.}
\end{figure}

The ratio of X-ray to bolometric luminosity is a good indicator of
stellar magnetic activity. For all sample stars, X-ray luminosity is
available in the NEXXUS database \citep[i.e., ROSAT data,][]{NEXXUS}.
Bolometric luminosity is calculated using the calibration in
\cite{Delfosse98} with $M_\mathrm{V}$ from \cite{Gliese91} and $R-I$
color from \cite{Leggett92}. The ratio of X-ray to bolometric
luminosity for the stars of this sample is plotted versus $v\,\sin{i}$
in the left panel of Fig.\,\ref{fig:LX_vsini}, and versus the Rossby
number (the ratio of Period to convective overturn time) in
Fig.\,\ref{fig:LX_Ro}.

It should be noted that X-ray luminosities from the ROSAT measurements
have very high internal precision, but there is some debate about the
conversion factor between the count rate and the flux. \cite{Schmitt95}
also reports X-ray luminosities for stars in the solar neighborhood
from ROSAT data using mostly the same data but a slightly different
conversion factor than the one in \cite{NEXXUS}.  Although the
differences between the two calibrations are sometimes significant,
they do not change the interpretation of the sample used here.  The
same is true for intrinsic variability of X-ray luminosities, which in
the active M dwarfs can be on the order of a few tenths of a dex
\cite[e.g.,][]{Robrade05}.

\subsubsection{Activity and projected rotation velocity}

The sample used here consists of two groups; I) stars with high X-ray
to bolometric luminosity ratio ($\log{L_{\mathrm X}/L_{\mathrm{bol}}}
> -3.5$) and II) stars with a low ratio of $\log{L_{\mathrm
    X}/L_{\mathrm{bol}}} < -4.5$. All stars in group II) have values
of $v\,\sin{i} \le 2$km\,s\,$^{-1}$ while the two stars in group I)
(AD~Leo and YZ~Cmi) are more rapid rotators. The two groups of stars
are clearly separated in the $L_{\mathrm X}/L_{\mathrm{bol}}$ vs.
$v\sin{i}$ plot (Fig.\,\ref{fig:LX_vsini}). The most obvious result
from this work is that the bulk of the very inactive stars (group II)
indeed has rotation velocities significantly lower than $v\,\sin{i} =
3$\,km\,s$^{-1}$.

Among the low-activity group, projected rotation velocities between 1
and 2\,km\,s$^{-1}$ are measured. The detected (projected) rotation
velocities are consistent with a rotation-activity connection in the
sense that most stars at $v\,\sin{i} \sim 2$\,km\,s$^{-1}$ have higher
$\log{L_{\mathrm X}/L_{\mathrm{bol}}}$ than the slowest rotators at
$v\,\sin{i} \sim 1$\,km\,s$^{-1}$. Seven of the eight objects (among
the low-activity subsample) follow a rather tight trend that falls
only slightly below the relation expected from an extrapolation of the
rotation-activity connection found in hotter stars (dashed line in
Fig.\,\ref{fig:LX_vsini}; note that Gl\,205 and Gl\,514 share the same
point). The only star that is a little off the track is Gl\,526, which
has the highest rotation velocity ($v\,\sin{i} = 2.0$\,km\,s$^{-1}$)
among the low-activity subsample, but even among them shows comparably
low activity ($\log{L_{\mathrm X}/L_{\mathrm{bol}}} < -5$).

It has to be taken into account that it is only the \emph{projected}
rotation velocity that is measured here, so that the true rotation
speed indeed is higher than $v\,\sin{i}$. Statistically, every
measurement is underestimated by a factor of $\pi/4$, but that does
not help intercomparing rotation velocities in a sample. The
projection may move the measured rotation velocities of individual
stars further to the right in the left panel of
Fig.\,\ref{fig:LX_vsini}, however, it cannot be expected that this has
significant influence on the trend among the seven stars mentioned.
Correcting for the projection effect in a statistical sense does
indeed move the stars further away from the dashed line, but as was
mentioned earlier, the absolute position of this line is not a crucial
point.

Considering Gl\,526 as an outlier, seven objects define a very close
rotation-activity connection in the early M-dwarfs rotating slower
than 3\,km\,s$^{-1}$. Gl\,526 has a spectral type of M\,1.5 and there
is no reason to believe that there is any difference to the other
seven slow rotators. A small inclination angle would mean that Gl\,526
is indeed rotating even more rapidly, so a projection effect would
make the inconsistency worse. However, the intrinsic scatter among
hotter stars in the rotation-activity relation is often much stronger
than the difference between Gl\,526 and the other stars of the sample.

The two sample stars with a much higher ratio of X-ray to bolometric
luminosity in the sample, AD~Leo and YZ~Cmi, are consistent with the
rotation-activity connection in the sense that they are more active
while rotating more rapidly than the inactive stars of the sample.
However, the relatively large difference of more than one order of
magnitude in $L_{\mathrm X}/L_{\mathrm{bol}}$ is only accompanied by a
difference of 1--2\,km\,s$^{-1}$ in $v\,\sin{i}$ between AD~Leo and
the slower rotators.  This would be a rather abrupt change in activity
or a very steep rise in the rotation-activity connection at velocities
around 3\,km\,s$^{-1}$.  \cite{SH86} report a rotation period of
AD~Leo of 2.7\,d which would indicate a higher rotation velocity
around $v\, \approx 10$\,km\,s$^{-1}$ observed under a rather low
inclination angle (see Section\,\ref{sect:ActRossby}).

At high rotation, there is no doubt that among early M-stars activity
saturates rather early (at comparably low rotation rates) while very
slow rotators ($v\,\sin{i} \la 2$\,km\,s$^{-1}$) have very low
activity ($\log{L_{\mathrm X}/L_{\mathrm{bol}}} \la -4.5$).  The gap
between the ``very slow'' rotators and the saturation regime could not
be bridged with this sample, and observations of stars probing
activity around $\log{L_{\mathrm X}/L_{\mathrm{bol}}} \approx -4$ is
required.

\subsubsection{Activity and Rossby number}
\label{sect:ActRossby}

\begin{figure}
  \includegraphics[width=.5\textwidth]{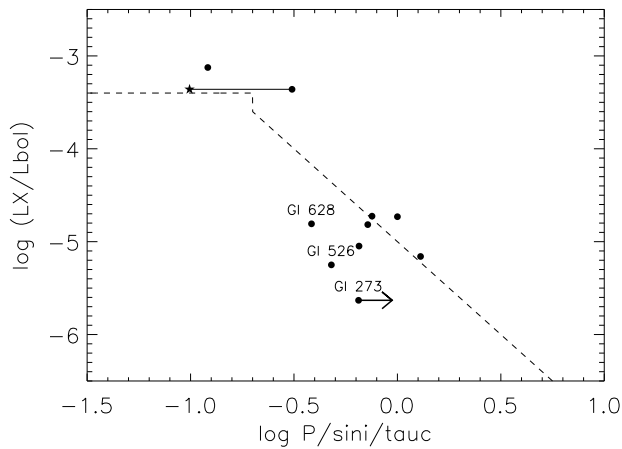}
  \caption{\label{fig:LX_Ro}Normalized X-ray luminosity against Rossby
    number for the sample. AD~Leo is plotted at two different Rossby
    numbers, the filled circle is the value from $v\,\sin{i}$
    measurements, the star is from the period given by \cite{SH86}.}
\end{figure}

So far the rotation-activity connection was only investigated in terms
of surface velocity. A parameter that seems to play a more important
role at least in solar-type stars is the Rossby number, that is the
ratio of the rotation period $P$ to the convective overturn time
$\tau_\mathrm{c}$ \citep{Noyes84}.  The ratio of X-ray to bolometric
luminosity is plotted against the Rossby number ($Ro$) in
Fig.\,\ref{fig:LX_Ro}.  To calculate $Ro$ for the sample object, the
period was derived using the projected rotation velocity and the radii
given in \cite{Lacy77}.  Thus, only the projected rotation period
$P/\sin{i}$ is available.  Convective overturn times for the sample
stars were calculated from Eq.~4 in \cite{Noyes84}. Other calculations
of \cite{Gilliland86} and more recently \cite{Kim96} differ only
little from the results of \cite{Noyes84}. Since the spectral types of
the sample are all very similar, using other versions of
$\tau_\mathrm{c}$ changes the results very little (in fact, it would
shift all points \emph{and} the relation expected from hotter stars
towards lower Rossby numbers).

The individual values of $\tau_\mathrm{conv}$ used for the calculation
of $Ro$ may be questioned.  Nevertheless, there is a clear indication
that the field M-dwarfs in Fig.\,\ref{fig:LX_Ro} generally follow the
same rotation-activity relation as hotter and younger stars. Stars
with the smallest Rossby numbers show saturated emission of
$\log{L_{\mathrm X}/L_{\mathrm{bol}}} \approx -3$, those with higher
Rossby number are less active. The scatter in Rossby number among the
inactive stars at given $L_\mathrm{X}/L_\mathrm{bol}$ is about half a
dex, which is not larger than the scatter in samples of hotter stars
on the non-saturated part of the rotation-activity connection
\citep[e.g.][]{Patten96}.  AD~Leo may have a somewhat high (projected)
Rossby number for its high activity. This may be due to a small
inclination angle as mentioned above.  The position of AD~Leo shifts
significantly to the left if the suggested period of 2.7~d
\citep{SH86} is used, it is indicated with a star instead of a circle
in Fig.\,\ref{fig:LX_Ro}. However, the presence of a rotation period
of 2.7~d of AD~Leo in the data of \cite{SH86} may be debatable,
\cite{Torres82} also report a period of 2.6\,d but they also put a
question mark to their result. On the other hand, regardless of what
the rotation period of AD~Leo really is, even the longest possible
period of 8.4~d taken from the new value of $v\,\sin{i}$ would lead to
a Rossby number smaller than the ones in the inactive stars.

Finally, the Rossby number at which activity becomes saturated is
consistent with the rotation-activity relation in hotter stars.
\cite{Patten96} shows that the transition from the active stars to
inactive stars occurs rather rapidly around a value of $Ro \approx
0.3$ ($\log{Ro} \approx -0.5$), just where it occurs in
Fig.\,\ref{fig:LX_Ro}, as well.

\section{Summary}
\label{sect:summary}

This study probes the validity of the rotation-activity connection --
i.e. stars show saturated activity at low Rossby numbers and
decreasing activity as Rossby number grows larger than $\sim 0.3$ --
in stars with very deep convection zones near the boundary to fully
convective stars. These early M-stars have very long convective
overturn times, so that stars populating the non-saturated part of the
rotation-activity connection need to be very slow rotators with
rotation periods of several weeks or equatorial rotation velocities
lower than $\sim 3$\,km\,s$^{-1}$. Accurate rotation velocities were
measured in eight probably not fully convective stars with low
activity ($\log{L_\mathrm{X}/L_\mathrm{bol}} < -4.5$) for which only
upper limits in $v\,\sin{i}$ were available (for Gl\,229A, a detection
of $v\,\sin{i} = 3$\,km\,s$^{-1}$ was reported earlier, but here
$v\,\sin{i} = 1$\,km\,s$^{-1}$ is measured). All eight inactive stars
show projected rotation velocities of or below $v\,\sin{i} =
2$\,km\,s$^{-1}$.

For the first time, narrow absorption lines of molecular FeH were
intrinsically resolved. Comparison to a model of an atomic Fe~I
absorption line shows that the line shape of that model is
qualitatively consistent with the observations. This allows a deeper
investigation of M-star convective patterns in high quality data in
this spectral range.

The main conclusion of this study is that the rotation activity
connection that is valid in stars of spectral types F--M among rapidly
rotating cluster stars, and stars of spectral type F--K in the field
\citep{Noyes84, Patten96, Pizzolato03}, is still valid in the early
field M-dwarfs after they have spun down to very low rotation rates.
Thus, the rotation activity connection holds in all main-sequence
field stars that harbor a convective envelope.

\acknowledgements{The author thanks Thomas Dall, who was the
  instrument scientist at the time of observation, for professional
  and very enjoyable support, and S. Wedemeyer and H.-G. Ludwig for
  providing the Fe line profile. AR has received research funding from
  the European Commission's Sixth Framework Programme as an Outgoing
  International Fellow (MOIF-CT-2004-002544).}

\end{document}